\documentclass[a4paper,11pt]{article}
\usepackage{graphicx,amssymb,bm,latexsym,color,epsf,cite}
\makeatletter
\@addtoreset{equation}{section}

\makeatother
\newcommand{\bea}{\begin{eqnarray}}
\newcommand{\eea}{\end{eqnarray}}
\newcommand{\be}{\begin{equation}}
\newcommand{\ee}{\end{equation}}
\newcommand{\vs}[1]{\vspace{#1 mm}}
\newcommand{\hs}[1]{\hspace{#1 mm}}
\renewcommand{\a}{\alpha}
\renewcommand{\b}{\beta}
\renewcommand{\c}{\gamma}
\newcommand{\G}{\Gamma}
\renewcommand{\d}{\delta}

\newcommand{\s}{\sigma}

\newcommand{\la}{\lambda}
\newcommand{\bomega}{\bar\omega}

\newcommand{\nn}{\nonumber\\}
\newcommand{\p}[1]{(\ref{#1})}

\newcommand{\fb}{\bar f}
\newcommand{\br}{\bar R}
\newcommand{\bR}{\bar R}
\newcommand{\bg}{\bar g}

\newcommand{\bnabla}{\bar\nabla}

\newcommand{\Det}{{\rm Det}}

\newcommand{\sdiff}{\mathit{SDiff}}
\newcommand{\lich}{{\Delta_L}}

\begin{document}

\begin{flushright}
KU-TP 072 \\
\today
\end{flushright}

\begin{center}
{\Large\bf $f(R, R_{\mu\nu}^2)$ at one loop}
\vs{10}

{\large
N. Ohta,$^{a,b,}$\footnote{e-mail address: ohtan@phys.kindai.ac.jp}
R. Percacci$^{c,d,}$\footnote{e-mail address: percacci@sissa.it}
and A. D. Pereira$^{e,}$\footnote{e-mail address: a.pereira@thphys.uni-heidelberg.de}
} \\
\vs{10}
$^a${\em Department of Physics, Kindai University,
Higashi-Osaka, Osaka 577-8502, Japan}

$^b${\em Maskawa Institute for Science and Culture,
Kyoto Sangyo University, Kyoto 603-8555, Japan}

$^c${\em International School for Advanced Studies, via Bonomea 265, 34136 Trieste, Italy}

$^d${\em INFN, Sezione di Trieste, Italy}

$^e${\em 
Institut f\"ur Theoretische Physik, Universit\"at Heidelberg,\\
Philosophenweg 12, 69120 Heidelberg, Germany}

\vs{10}
%%%%%%%%%%%%%%%%%%%%%%%%%%%%%%%%
{\bf Abstract}
\end{center}
We compute the one-loop divergences in a theory of gravity
with Lagrangian of the general form $f(R,R_{\mu\nu}R^{\mu\nu})$,
on an Einstein background.
We also establish that the one-loop effective action 
is invariant under a duality
that consists of changing certain parameters in the
relation between the metric and the quantum fluctuation field. 
Finally, we discuss the unimodular version of such a theory and establish its equivalence
at one-loop order with the general case.

%%%%%%%%%%%%%%%%%%%%%%%%%%%%
\section{Introduction}
%%%%%%%%%%%%%%%%%%%%%%%%%%%%

The calculation of the one-loop divergences in Einstein theory
without cosmological constant
was performed originally in \cite{tHooft:1974toh}
and established the perturbative nonrenormalizability of the theory
in the presence of matter.
The cosmological constant was included in \cite{duff}.
The nonrenormalizability of pure gravity requires a
two-loop calculation that was only done much later \cite{Goroff:1985th}.
One-loop divergences in higher-derivative gravity
(containing terms quadratic in the Ricci scalar, Ricci tensor
and Riemann tensor)
were calculated originally in \cite{julve,ft1,avrabar}.
The theory is renormalizable \cite{stelle}
and one obtains beta functions
that in a certain regime lead to asymptotic freedom.

This calculation was redone and extended to $4+\epsilon$
dimensions in \cite{deBerredoPeixoto:2004if}.
In a context of asymptotic safety, these beta functions were 
reproduced starting from the functional renormalization group
in \cite{Codello:2006in,niedermaier}
and in other dimensions in \cite{OP2013},
and pushed beyond the one-loop approximation in 
\cite{lauscher, bms,sgrz}.
The conformal case requires a separate calculation
\cite{deBerredoPeixoto:2003pj,OP2016}.
Higher derivative gravity has been revisited recently
\cite{AKKLR}.
The main issue with this theory is the presence of ghosts
in the perturbative spectrum.
Over time, several ways out have been suggested
\cite{salam,Tomboulis,floper4,bonannoreuter1}, and more recently
\cite{Mannheim:2006rd,smilga,Salvio:2014soa,Einhorn:2014gfa,holdom,donhdg,Becker:2017tcx},
but this point remains unsettled, for the time being. 

Beyond the purely quadratic terms,
systematic investigations have been done only in the case
of $f(R)$ actions.
The one-loop divergences for such a theory,
on a maximally symmetric background have been computed
originally in \cite{Cognola:2005de}.
This calculation has been extended recently 
to arbitrary backgrounds \cite{RS}
and to Einstein spaces \cite{Ohta}.
Functions of the Ricci scalar only
can be recast as Einstein theory coupled to a scalar,
so in these theories one is not probing the effects due to
higher-derivative spin-two propagators.
In the functional renormalization group approach,
$f(R)$ gravity has been studied in
\cite{cpr1,ms,cpr2,fallslitim,benedetti,dm1,dm2,dsz,opv,fallsohta,CFPR,Alkofer:2018fxj} and in \cite{Eichhorn:2015bna}
in the unimodular setting. 
The classically equivalent scalar-tensor theories
have been studied for example in \cite{pv1}.
Recently there has also been work on Lagrangians of the form
$f(R_{\mu\nu}R^{\mu\nu})+R Z(R_{\mu\nu}R^{\mu\nu})$
\cite{Falls:2017lst}.

In this paper we will consider more general theories
depending on the Ricci scalar and the square of the Ricci tensor
\be
S(g)=\int d^dx\sqrt{|g|}\,f(R,X)\ ,
\ee
where
$$
X=R_{\mu\nu}R^{\mu\nu}\ .
$$
Our treatment will be limited to Einstein backgrounds,
where some powerful tools allow us to reduce the
calculation to that of determinants of second order operators only.

An important aspect of these calculations is that they
only give universal results on shell.
By ``universal" we mean independent of arbitrary choices such as the choice of gauge and
parametrization of the quantum field.
The off-shell dependence of one-loop divergences
on the choice of gauge and parametrization has been
investigated in \cite{kallosh,Kalmykov:1995fd,KP};
see also \cite{Goncalves:2017jxq} for a recent account.
In the context of asymptotic safety, this dependence
has been explored in \cite{Gies:2015tca,nink}.
More recently, we have computed the gauge and parametrization
dependence of the one-loop divergences
in Einstein gravity \cite{oppI}
and higher-derivative gravity \cite{oppII}
(with four free parameters altogether).
We will use this general parametrization also in this paper.

One curious feature of the result is its invariance under
a certain discrete, idempotent change of the parameters,
that involves the replacement of a densitized metric by
a densitized inverse metric, that we called ``duality.''
The existence of this invariance was first noticed in
\cite{oppI} in the case of Einstein gravity and
in \cite{oppII} in higher-derivative gravity on an Einstein space.
The question whether it exists also in more general context
was one of the original motivations for this work.
We find that the answer is positive.

This paper is organized as follows.
The long Sec.~2 contains most of the technical steps,
including the derivation of the second variation of the
action in terms of York variables and the gauge fixing.
In Sec.~3 the main results are presented and conclusions drawn.
Section 4 contains the discussion of the unimodular case.

%%%%%%%%%%%%%%%%%%%%%%%%%%%%%%%%%%%%%%%%%%%%%%%%%%%
\section{The Hessian}
%%%%%%%%%%%%%%%%%%%%%%%%%%%%%%%%%%%%%%%%%%%%%%%%%%%

\subsection{Variations}

For a one-loop calculation we need to expand the action
to second order in the fluctuation field.
We begin by considering the linear splitting
\be
g_{\mu\nu}=\bg_{\mu\nu}+h_{\mu\nu}\ .
\ee
The function $f(R, R_{\mu\nu}^2)$ can be expanded as
\bea
f(R, R_{\mu\nu}^2) \!\! &=&\!\! \fb + \fb_R (R^{(1)} + R^{(2)})+ \frac12 \bar f_{RR} (R^{(1)})^2
\nn \!\! &&\!\! + \;
\bar f_{X} (X^{(1)} + X^{(2)})
+\frac12 \bar f_{XX} (X^{(1)})^2 + \bar f_{RX} R^{(1)} X^{(1)}
+ \cdots,
\eea
where the subscripts on $f$ denote derivatives with respect to its arguments:
$$
f_R=\frac{\partial f}{\partial R}\ ,\qquad
f_X=\frac{\partial f}{\partial X}\ ,\qquad
f_{RR}=\frac{\partial^2 f}{\partial R^2}\ ,\qquad
f_{RX}=\frac{\partial^2 f}{\partial R\partial X}\ ,\qquad
f_{XX}=\frac{\partial^2 f}{\partial X^2}\ .
$$
The bar on any quantity means that it is evaluated on the
background [for example, $\fb_X=f_X(\bR,\bar X)$]
and we denote by a superscript in parentheses the power of $h$ contained
in a certain term of the expansion.
Thus we have 
\bea
R \!\! &=&\!\! \br + R^{(1)} + R^{(2)} + \cdots,\nn
R_{\mu\nu}R^{\mu\nu}  \equiv X\!\! &=& \!\!\bar X + X^{(1)} + X^{(2)} + \cdots.
\eea
Explicitly we have for the Ricci tensor
\bea
R^{(1)}_{\mu\nu}\!\! &=&\!\! -\frac12 (\bnabla_\mu \bnabla_\nu h
- \bnabla_\mu h_{\nu} - \bnabla_\nu h_\mu + \Box h_{\mu\nu})
- \br_{\a\mu\b\nu}h^{\a\b}+\frac12 \br_{\mu\a}h^\a_\nu +\frac12 \br_{\nu\a} h^\a_\mu
, \nn
R^{(2)}_{\mu\nu}\!\! &=&\!\! \frac12 \bnabla_\mu(h^{\a\b} \bnabla_\nu h_{\a\b})
-\frac12 \bnabla_\a \{h^{\a\b}( \bnabla_\mu h_{\nu\b}+ \bnabla_\nu h_{\mu\b}
- \bnabla_\b h_{\mu\nu}) \} \nn
&& \hs{-10}
- \frac14 (\bnabla_\mu h^\b_\a+ \bnabla_\a h^\b_\mu -\bnabla^\b h_{\a\mu})
(\bnabla_\b h^\a_\nu + \bnabla_\nu h^\a_\b -\bnabla^\a h_{\b\nu})
+ \frac14 \bnabla_\a h (\bnabla_\mu h^\a_\nu + \bnabla_\nu h_\mu^\a -\bnabla^\a h_{\mu\nu}),\nn
\label{ricexp}
\eea
where we have used the notation $h_\mu= \bnabla^\nu h_{\mu\nu}$ and $\Box = \bnabla^{2}$.
This implies
%From also have~\cite{OP2013}
\bea
X^{(1)}(h_{\mu\nu}) \!\! &=&\!\! 2(\br^{\mu\nu} R^{(1)}_{\mu\nu} - \br_{\mu\nu}\br^\mu{}_\rho h^{\nu\rho}), \nn
X^{(2)}(h_{\mu\nu}) \!\! &=&\!\! (R^{(1)}_{\mu\nu})^2 +2 \br^{\mu\nu} R^{(2)}_{\mu\nu}
-4 \br^{\mu\nu} h_\nu{}^\rho R^{(1)}_{\mu\rho}
+2 \br_{\mu\la} \br_\nu{}^\la (h^2)^{\mu\nu} +\br_{\mu\nu} \br_{\la\rho}h^{\mu\la} h^{\nu\rho}.
\eea
For the Ricci scalar we have
%Explicitly for the linear parametrization, they are given by~\cite{OP2013}
\bea
R^{(1)}(h_{\mu\nu}) \!\! &=&\!\! \bnabla_\mu h^\mu -\Box h - \br_{\mu\nu} h^{\mu\nu}, \nn
R^{(2)}(h_{\mu\nu}) \!\! &=&\!\! \frac34 \bnabla_\a h_{\mu\nu} \bnabla^\a h^{\mu\nu} +h_{\mu\nu} \Box h^{\mu\nu}
-h_\mu^2 + h_\mu \bnabla^\mu h -2 h_{\mu\nu} \bnabla^\mu h^\nu
+h_{\mu\nu} \bnabla^\mu \bnabla^\nu h \nn
&& \hs{-2}
-\; \frac12 \bnabla_\mu h_{\nu\a} \bnabla^\a h^{\mu\nu} -\frac14 \bnabla_\mu h \bnabla^\mu h
+\br_{\a\b\c\d} h^{\a\c} h^{\b\d}.
\label{rsexp}
\eea

We will assume that the background is Einstein
\bea
\br_{\mu\nu} = \frac{\br}{d} \bg_{\mu\nu}\ .
\label{Einstein}
\eea
Then, the covariant derivatives of the Ricci tensor vanish
and the Ricci scalar is constant. 
When $R^{(2)}$ appears linearly in the expansion of the action, 
we can make partial integration and use a simpler expression:
\bea
R^{(2)}(h_{\mu\nu}) = \frac14 (h_{\mu\nu} \Box h^{\mu\nu} +h\Box h +2 h_\mu^2
+2 \br_{\mu\nu} h^{\mu\a} h^\nu_\a +2 \br_{\a\b\c\d} h^{\a\c} h^{\b\d}).
\eea
For the same reason, when $R^{(2)}_{\mu\nu}$ 
appears in the action linearly, 
we can neglect the first line of (\ref{ricexp}).

The last ingredient for the linear expansion is the expansion
of the measure
\bea
\sqrt{|g|} = \sqrt{|\bg|}\left(1+\frac12 h + \frac{h^2-2h_{\mu\nu}^2}{8}+ \cdots \right),
\eea

As in \cite{oppI,oppII}, we will calculate the Hessian for
a more general parametrization of the fluctuations, namely
\be
\label{gexp}
g_{\mu\nu}=\bg_{\mu\nu}+\delta g_{\mu\nu}\ ,
\ee
where the fluctuation is expanded:
\be
\label{gammaexp}
\delta g_{\mu\nu}=
\delta g^{(1)}_{\mu\nu}
+\delta g^{(2)}_{\mu\nu}
+\delta g^{(3)}_{\mu\nu}+\ldots \ ,
\ee
where $\delta g^{(n)}_{\mu\nu}$ contains $n$
powers of $h_{\mu\nu}$.
We will parametrize the first two terms of the expansion
as follows:
\bea
\delta g^{(1)}_{\mu\nu}&=&h_{\mu\nu}+m\bg_{\mu\nu}h \ ,
\nonumber\\
\delta g^{(2)}_{\mu\nu}&=&
\omega h_{\mu\rho}h^\rho{}_\nu
+m h h_{\mu\nu}
+m\left(\omega-\frac{1}{2}\right)\bg_{\mu\nu}h^{\alpha\beta}h_{\alpha\beta}
+\frac{1}{2}m^2\bg_{\mu\nu}h^2\ .
\label{deltag}
\eea
As explained in \cite{oppI},
the parameter $m$ is related to the weight 
(in the sense of tensor density)
of the full quantum field
\be
\gamma_{\mu\nu}=g_{\mu\nu}(\det(g_{\mu\nu}))^{-\frac{m}{1+dm}}
\ ,\qquad
\gamma^{\mu\nu}=g^{\mu\nu}(\det(g_{\mu\nu}))^{\frac{m}{1+dm}}\ .
\ee
We consider four main types expansion:
linear expansion of the densitized metric or its inverse
\be
\label{linexp}
\gamma_{\mu\nu}=\bar\gamma_{\mu\nu}+\hat h_{\mu\nu}
\qquad \mathrm{or}\qquad
\gamma^{\mu\nu}=\bar\gamma^{\mu\nu}-\hat h^{\mu\nu}\ ,
\ee
and exponential expansion of the densitized metric or its inverse
\be
\label{expexp}
\gamma_{\mu\nu}=\bar\gamma_{\mu\rho}(e^{\hat h})^\rho{}_\nu
\qquad \mathrm{or}\qquad
\gamma^{\mu\nu}=(e^{-\hat h})^\mu{}_\rho\bar\gamma^{\rho\nu}\ .
\ee
The parameter $\omega$ discriminates among
the linear expansion of the covariant metric (for $\omega=0$);
the exponential expansion (for $\omega=1/2$,
independent of whether one expands the metric or its inverse);
and the linear expansion of the inverse metric (for $\omega=1$).

We found in \cite{oppI,oppII} that the divergences 
of Einstein theory and of higher-derivative gravity
on an Einstein background are
invariant under the following change of the parameters:
\bea
(\omega,m)\mapsto\left(1-\omega,-m-\frac{2}{d}\right)\ .
\label{duality}
\eea
Since this maps the expansion of the metric to the
expansion of the inverse metric,
we referred to it as duality.
We also observed that the case $\omega=1/2$, $m=-1/d$,
which corresponds to the exponential expansion
of unimodular gravity,
is a fixed point of the duality and is a point of minimum
sensitivity for the parameter dependence of the 
off-shell divergences.

We are going to use again the parametrization defined above
and check whether the results are duality invariant
in the case of the theory $f(R,X)$.
To this end we insert the
preceding parametrization in the second order linear expansion
of the metric and we obtain
\bea
&& \bar f_R \left(\left. R^{(1)}\right|_{h_{\mu\nu}\to \d g^{(2)}_{\mu\nu}}
 + \left.R^{(2)}\right|_{h_{\mu\nu}\to \d g^{(1)}_{\mu\nu}} \right)
+ \frac12 \bar f_{RR} \left(\left. R^{(1)}\right|_{h_{\mu\nu}\to \d g^{(1)}_{\mu\nu}}\right)^2
\nn && +\;
\bar f_{X} \left(\left. X^{(1)}\right|_{h_{\mu\nu}\to \d g^{(2)}_{\mu\nu}}
+\left. X^{(2)}\right|_{h_{\mu\nu}\to \d g^{(1)}_{\mu\nu}}\right)
+ \frac12 \bar f_{XX} \left(\left. X^{(1)}\right|_{h_{\mu\nu}\to \d g^{(1)}_{\mu\nu}}\right)^2
\nn && +\;
\bar f_{RX} \left.\left( R^{(1)} X^{(1)}\right)\right|_{h_{\mu\nu}\to \d g^{(1)}_{\mu\nu}}
+ \frac12 \fb \d g^{(2)} +\frac18  \fb\left( (\d g^{(1)})^2 - 2(\d g_{\mu\nu}^{(1)})^2 \right)
\nn && +\,
\left. \frac{\d g^{(1)}}{2} \left( \bar f_R R^{(1)} +\bar f_{X} X^{(1)}\right)
\right|_{h_{\mu\nu}\to \d g^{(1)}_{\mu\nu}} \nn
&=& \frac18  \fb\left( 4 \d g^{(2)}+ (\d g^{(1)})^2 - 2(\d g_{\mu\nu}^{(1)})^2 \right)
+ \bar f_R \left( R^{(1)} (\d g^{(2)}_{\mu\nu})
+ R^{(2)}(\d g^{(1)}_{\mu\nu}) + \frac{\d g^{(1)}}{2} R^{(1)}(\d g^{(1)}_{\mu\nu}) \right) \nn
&& +\; \frac12 \bar f_{RR} \left( R^{(1)}(\d g^{(1)}_{\mu\nu})\right)^2
+ \bar f_{X} \left( X^{(1)} (\d g^{(2)}_{\mu\nu})
+ X^{(2)}(\d g^{(1)}_{\mu\nu}) + \frac{\d g^{(1)}}{2} X^{(1)}(\d g^{(1)}_{\mu\nu}) \right) \nn
&& +\; \frac12 \bar f_{XX} \left( X^{(1)}(\d g^{(1)}_{\mu\nu})\right)^2
+ \bar f_{RX} R^{(1)}(\d g^{(1)}_{\mu\nu}) X^{(1)}(\d g^{(1)}_{\mu\nu}),
\eea
where $\d g^{(1)}=\bg^{\mu\nu} \d g^{(1)}_{\mu\nu}$ and $\d g^{(2)}=\bg^{\mu\nu} \d g^{(2)}_{\mu\nu}$.

%%%%%%%%%%%%%%%%%%%%%%%%%%%%%%%
\subsection{York decomposition}

The structure of the Hessian on an Einstein space is  
much clearer when one separates the spin two, one and zero
components of the fluctuation field:
\bea
h_{\mu\nu} = h^{TT}_{\mu\nu} + \bnabla_\mu\xi_\nu + \bnabla_\nu\xi_\mu +
\bnabla_\mu \bnabla_\nu \s -\frac{1}{d} \bg_{\mu\nu} \bnabla^2 \s +
\frac{1}{d} \bg_{\mu\nu} h,
\label{york}
\eea
where
$$
\bnabla^\mu h^{TT}_{\mu\nu} = 0\ ;
\qquad
\bg^{\mu\nu} h^{TT}_{\mu\nu}=0\ ;\qquad
 \bnabla_\mu \xi^\mu=0\ .
$$
In the path integral this change of variables gives rise
to a Jacobian that will be discussed later.
When \p{york} is squared, we get
\bea
\int d^d x \sqrt{|\bg|}\Big[ h^{TT}_{\mu\nu}h^{TT\,\mu\nu} +2 \xi_\mu \Big(\Delta_{L1}
-\frac{2}{d}\br \Big)\xi^\mu +\frac{d-1}{d}\s \Delta_{L0}\Big( \Delta_{L0}-\frac{\br}{d-1} \Big)\s
+\frac{1}{d} h^2 \Big].
\label{f1}
\eea
where $\Delta_{L2}$, $\Delta_{L1}$ and $\Delta_{L0}$
are the Lichnerowicz Laplacians defined as
\bea
\Delta_{L2} T_{\mu\nu} &=& -\bnabla^2 T_{\mu\nu} +\br_\mu{}^\rho T_{\rho\nu}
+ \br_\nu{}^\rho T_{\mu\rho} -\br_{\mu\rho\nu\s} T^{\rho\s} -\br_{\mu\rho\nu\s} T^{\s\rho}, \nn
\Delta_{L1} V_\mu &=& -\bnabla^2 V_\mu + \br_\mu{}^\rho V_\rho, \nn
\Delta_{L0} S &=& -\bnabla^2 S.
\eea
%on the symmetric tensor, vector and scalar respectively.
Note that we can freely insert the covariant derivatives inside the above expression.
We also have the useful formulae
\bea
\bnabla_\mu h^\mu_\nu &=&- \left(\lich_1 -\frac{2}{d}\br \right) \xi_\nu
 -\frac{d-1}{d} \bnabla_\nu \left(\lich_0 -\frac{\br}{d-1} \right)\s +\frac{1}{d} \bnabla_\nu h, \nn
\bnabla_\mu \bnabla_\nu h^{\mu\nu} &=& \frac{d-1}{d} \lich_0 \left(\lich_0 -\frac{\br}{d-1} \right)\s
-\frac{1}{d} \lich_0 h.
\eea

We then find that the Hessian is
\footnote{From here on we always consider the Euclidean case.}
\bea
S^{(2)}=
\int d^dx\sqrt{\bg}
\left[
h^{TT}_{\mu\nu}H^{TT}h^{TT\mu\nu}
+\xi_\mu H^{\xi\xi} \xi^\mu
+\sigma H^{\sigma\sigma} \sigma
+\sigma H^{\sigma h} h
+h H^{h\sigma} \sigma
+h H^{hh} h
\right],
\eea
where
\bea
H^{TT} &=& \frac14 \left[ \left\{\bar f_X \left( \lich_2 -\frac{4\br}{d} \right)-\bar f_R\right\}
\left( \lich_2 -\frac{2\br}{d}\right) - (1-2\omega)(1+md) \tilde E \right], \\
H^{\xi\xi} &=& - \frac{(1-2\omega)(1+md)}{2} \left(\lich_1 -\frac{2\br}{d}\right) \tilde E, \\
H^{\sigma\sigma} &=& \frac12 \left(\frac{d-1}{d}\right)^2 
\left[ 
P \lich_0 \left(\lich_0 -\frac{\br}{d-1} \right) 
 +Q \lich_0
 \right. \nn
&& \left. \
-\frac{d(1-2\omega)(1+md)}{2(d-1)} \tilde E \right] 
%\nn
%&& \hs{30} \times\, 
\lich_0 \left(\lich_0-\frac{\br}{d-1}\right), \\ 
H^{\sigma h} &=& \left(\frac{d-1}{d}\right)^2 \frac{1+md}{2} \left[P \left(\lich_0 -\frac{\br}{d-1}\right)
+Q \right]
 \lich_0 \left(\lich_0 -\frac{\br}{d-1}\right), \\
H^{hh} &=& \left(\frac{d-1}{d}\right)^2 \frac{(1+md)^2}{2} 
\left[ P \left(\lich_0 - \frac{\br}{d-1} \right)^2 
%&& 
+\, Q\left(\lich_0 -\frac{\br}{d-1}\right)
\right.
\nn
&&
\left.
\qquad\qquad\qquad\qquad\qquad\qquad
+ \frac{d[(1+md)d-2(1-2\omega)]}{4(d-1)^2 (1+md)} \tilde E \right] ,
\eea
where we used the shorthands
\bea
P&=&\bar f_{RR}+\frac{4}{d^2}\br^2 \bar f_{XX}+4\br \bar f_{RX}
+ \frac{d}{2(d-1)} \bar f_X  
\\
Q&=&\frac{d-2}{2(d-1)} \bar f_R + \frac{3d^2-10d+8}{2d(d-1)^2}\br \bar f_X
\eea
and
\bea
\tilde E \equiv \bar f - \frac{2}{d}\br \bar f_R -\frac{4\br^2}{d^2} \bar f_X =0,
\label{onshell2}
\eea
is the field equation evaluated on the Einstein space 
\p{Einstein}.

%%%%%%%%%%%%%%%%%%%%%%%%%%%
\subsection{Gauge fixing}

We now use a trick that has proven convenient in the case of general relativity (GR) and more generally of
$f(R)$ theories. A cleaner separation of physical and gauge degrees of freedom can be achieved by changing
variables in the scalar sector. Instead of $\sigma$ and $h$ we define:
\bea
\label{invscalar}
s &=& \lich_0 \sigma+(1+dm)h, \\
\chi &=& \sigma+\frac{b}{(d-1-b)\lich_0-\br}s 
= \frac{(d-1)\lich_0-\br}{(d-1-b)\lich_0-\br}\sigma
+\frac{b(1+dm)}{(d-1-b)\lich_0-\br}h\ ,
\label{chi}
\eea
where $s$ is gauge invariant and $b=\bar b/(1+md)$
is a gauge parameter that will be defined below.
A short calculation shows that the Jacobian of the transformation $(\s,h)\to(s,\chi)$ is 1.
It is easy to see that the whole scalar part of the Hessian
can be rewritten in the simple form
$\int d^dx\sqrt{\bg}\,s H^{ss} s$, where, on shell,
\bea
H^{ss}&=&\frac{1}{2}\left(\frac{d-1}{d}\right)^2
\left[P \left(\lich_0-\frac{\br}{d-1}\right)+Q\right]
\left(\lich_0-\frac{\br}{d-1}\right) .
\eea

Next we consider the gauge-fixing term
\bea
\label{gfaction}
S_{GF}=\frac{1}{2a}\int d^d x \sqrt{\bg}\,\bg^{\mu\nu}F_\mu F_\nu,
\eea
with
\bea
\label{gf}
F_{\mu}=\bar{\nabla}_{\alpha}{h^{\alpha}}_{\mu}-\frac{\bar b+1}{d}\bar{\nabla}_{\mu}h\,,
\eea
and $a$ and $\bar b$ are gauge parameters.
In terms of the new variables, the gauge-fixing action is
\bea
\label{gf2}
S_{GF}= - \frac{1}{2a}\int d^dx\sqrt{\bg}
\left[
\xi_\mu\left(\lich_1-\frac{2\br}{d}\right)^2\xi^\mu
+\frac{(d-1-b)^2}{d^2}
\chi\; \lich_0\left(\lich_0-\frac{\br}{d-1-b}\right)^2\chi
\right].~~
\eea
We see that on shell there is a perfect separation of
gauge and physical degrees of freedom:
the Hessian of the action only depends on $h_{\mu\nu}^{TT}$
and $s$, whereas the gauge-fixing term only depends on
$\xi_\mu$ and $\chi$.

As seen in \cite{oppII}, the use of the new variables 
also makes the gauge-fixing sector manifestly
invariant under the duality.

The ghost action contains a nonminimal operator
\bea
S_{gh}=i \int d^dx\sqrt{\bg}\,\bar C^\mu\left(
\delta_\mu^\nu\bnabla^2
+\left(1-2\frac{b+1}{d}\right)\bnabla_\mu\bnabla^\nu+\br_\mu{}^\nu\right)C_\nu .
\eea
Let us decompose the ghost into transverse and longitudinal parts
\bea
C_\nu=C^T_\nu+\bnabla_\nu C^L
=C^T_\nu+\bnabla_\nu\frac{1}{\sqrt{\lich_{0}}}C'^L ,
\eea
and the same for $\bar C$.
This change of variables has unit Jacobian.
Then, the ghost action splits into two terms:
\bea
\label{ghostaction}
S_{gh}= i \int d^dx\sqrt{\bg}
\left[ \bar C^{T\mu}\left(\lich_{1}-\frac{2\br}{d}\right)C^T_\mu
+2\frac{d-1-b}{d}
\bar C'^L\left(\lich_{0}-\frac{\br}{d-1-b}\right)C'^L\right].
\eea
We are now ready to calculate the one-loop divergences.

%%%%%%%%%%%%%%%%%%%%%%%%%%%%%%%%%%%
\section{Results}
%%%%%%%%%%%%%%%%%%%%%%%%%%%%%%%%%%%

%%%%%%%%%%%%%%%%%%%%%%%%%%
\subsection{Duality and parametrization independence}

These properties can be established already at the
level of the Hessian, which has been given in Sec.~2.2.
In the Hessian the parameters $\omega$ and $m$ appear only
in the combinations $1-2\omega$ and $1+md$.
These combinations change sign under the duality transformation~\p{duality}.
In $H^{TT}$, $H^{\xi\xi}$ and $H^{\sigma\sigma}$,
they appear in the combination
$(1-2\omega)(1+md)$, which is invariant.
Next we observe that in the scalar sector the two
terms on the diagonal are invariant and the two
off-diagonal terms change sign.
Thus the determinant and the trace of the Hessian
are both invariant, and so are its eigenvalues.
This is all that matters for the quantum theory.

We observe that all dependence on $\omega$ 
is proportional to the equation of motion.
The determinant has an on-shell dependence on $m$,
but only through an overall prefactor $(1+md)^2$,
which can be absorbed in a redefinition of $h$.
Thus all dependence on the parameters $m$ and $\omega$
goes away on shell, as it must.

%%%%%%%%%%%%%%%%%%%%%%%%%%%%%%%%%%%%%%
\subsection{One-loop, on-shell effective action}

Unless we (1) set $\omega=\frac12$, 
or (2) $m=-\frac{1}{d}$ or (3) go on shell, 
the effective action is gauge dependent.
We choose to impose the on-shell condition~\p{onshell2}. 
As we have already mentioned,
the result is then independent of $\omega$ and $m$.

The one-loop effective action is given by the
determinants coming from different components
of the fluctuation, and the ghosts and Jacobians.
%After taking the partial gauge fixing $h=0$,
The contributions are
\bea
\Det\left(\lich_2-\frac{4\br}{d}-\frac{\bar f_R}{\bar f_X} \right)^{-1/2}
\Det\left(\lich_2-\frac{2\br}{d} \right)^{-1/2} ,
\label{conTT}
\eea
from $h_{\mu\nu}^{TT}$,
\bea
\Det \left(\lich_1-\frac{2\br}{d}\right)^{-1} ,
\label{conxi}
\eea
from $\xi_\mu$,
\bea
\Det \left(\lich_0-\frac{\br}{d-1}\right)^{-1/2}
\Det \left( \lich_0 -\frac{\br}{d-1}
+ \frac{Q}{P}\right)^{-1/2} ,
\eea
from $s$,
\bea
\Det \lich_0^{-1/2} \Det\left(\lich_0-\frac{\br}{d-1-b}\right)^{-1} ,
\eea
from $\chi$, and finally
\bea
\Det \left(\lich_1-\frac{2\br}{d}\right)
\Det \left(\lich_0-\frac{\br}{d-1-b}\right) ,
\eea
from the ghost. In addition the York decomposition has Jacobian
\bea
{\Det} \Big(\lich_{1}-\frac{2}{d}\br \Big)^{1/2}
\Det [\lich_0]^{1/2}
{\Det} \Big(\lich_{0}-\frac{\br}{d-1} \Big)^{1/2}\ .
\label{conjac}
\eea

The effective action is then given by
\bea
\G \hs{-2}&=&\hs{-2} \frac12 \log\Det\left(\lich_2-\frac{4\br}{d} -\frac{\bar f_R}{\bar f_X} \right)
+ \frac12 \log\Det\left(\lich_2-\frac{2\br}{d} \right)
 \nn
&& \hs{-5}
+\frac12\log\Det\left(\lich_0-\frac{\br}{d-1}+\frac{Q}{P}\right)
-\frac12\log\Det\left(\lich_1-\frac{2\br}{d} \right)
.
\label{ea}
\eea
If $\bar f_X=0$, the first contribution is absent.
The gauge independence of the result is manifest.

%%%%%%%%%%%%%%%%%%%%%%%%%%%%%%%%%%%%%%%
\subsection{The logarithmic divergence}
%%%%%%%%%%%%%%%%%%%%%%%%%%%%%%%%%%%%%%%

The divergent part of the effective action
can be computed by standard heat kernel methods.
On an Einstein background in four dimensions, the logarithmically divergent part is
\bea
\Gamma_{log}(\bg) &=& \frac{1}{720(4\pi)^2}
\int d^4x\,\sqrt{\bg}
\log\left(\frac{\Lambda^2}{\mu^2}\right)
\left[
-826 \br_{\mu\nu\rho\sigma}^2
+509 \br^2 
- \frac{300 \br \bar f_R}{\bar f_X} 
- \frac{900 \bar f_R^2}{\bar f_X^2} \right. 
\nn
&&
\left.
+\frac{240\bR(3\bar f_R+2\bR \bar f_X)}{8\bar f_X+12\bar f_{RR}+48\bR \bar f_{RX}+3\bR^2 \bar f_{XX}}
-\frac{320(3\bar f_R+2\bR \bar f_X)^2}{(8\bar f_X+12\bar f_{RR}+48\bR \bar f_{RX}+3\bR^2 \bar f_{XX})^2}
\right], \nn
\label{gammaabc}
\eea
where $\Lambda$ stands for a cutoff and we introduced a reference mass scale $\mu$.
Note that on an Einstein space $\bar X=\bR^2/4$,
so in this formula $\bar f_R(\bR,\bar X)$, $\bar f_X(\bR,\bar X)$ etc, 
have to be interpreted as $\bar f_R(\bR,\bR^2/4)$, $\bar f_X(\bR,\bR^2/4)$ etc.,
so all the terms except the first are functions of $\bR$ only.

We can check this expression against two existing results
in the literature.
If we put
$$
f(R,X)=\alpha R^2+\beta X \ ,
$$
it reduces to
\be
\Gamma_{log}(\bg) = \frac{1}{(4\pi)^2}
\int d^4x\,\sqrt{\bg}
\log\left(\frac{\Lambda^2}{\mu^2}\right)
\left[
-\frac{413}{360} \br_{\mu\nu\rho\sigma}^2
-\frac{1200 \alpha ^2+200 \alpha  \beta -183 \beta ^2}{240 \beta ^2}\bR^2 \right] ,
\label{gammaabc1}
\ee
which is the standard universal result in higher derivative gravity.

On the other hand if we put 
$$
f(R,X)=f(R) \ ,
$$
we obtain
\be
\Gamma_{log}(\bg) = \frac{1}{(4\pi)^2}
\int d^4x\,\sqrt{\bg}
\log\left(\frac{\Lambda^2}{\mu^2}\right)
\left[
-\frac{71}{120} \br_{\mu\nu\rho\sigma}^2
+\frac{433}{1440}\bR^2
+\frac{\bar f_R \bR}{12\bar f_{RR}} 
-\frac{\bar f_R^2}{36\bar f_{RR}^2}\right],
\label{gammaabc2}
\ee
which agrees with the recent results of \cite{RS,Ohta}.

%%%%%%%%%%%%%%%%%%%%%%%%%%%%%%%%%%%
\section{Unimodular version}
%%%%%%%%%%%%%%%%%%%%%%%%%%%%%%%%%%%

It is of some interest to consider the unimodular version
of the same theory, by which we mean the theory with the same
Lagrangian density but with the metric constrained to satisfy
$$
\sqrt{|g|}=\bomega\ ,
$$
where $\bomega$ is some fixed scalar density
(usually chosen to be 1 in a class of coordinate systems).
This condition breaks the diffeomorphism group down to
the subgroup $\sdiff$ of ``special'' diffeomorphisms,
whose infinitesimal generators are transverse vector fields
$\epsilon^T_\mu$, satisfying
$$
\nabla^\mu \epsilon^T_\mu=0\ .
$$
The action is then
\be
S(g)=\int d^dx\,\bomega\,f(R,X)\ .
\ee
and in the variations one has to keep the determinant of $g$ fixed.
The simplest way of achieving this is to parametrize
the metric exponentially as in (\ref{expexp})
and to assume that the fluctuation field $h_{\mu\nu}$
is traceless
$$
\bg^{\mu\nu}h_{\mu\nu}=0\ .
$$
The second order expansion of the action is then identical
to the one in Sec.~2, except for the absence of all
the terms containing $h$ and for setting the parameter $\omega=1/2$.
The final Hessian is very simple and has only two contributions:
\bea
H^{TT} &=& \frac14\left[\bar f_X \left( \lich_2 -\frac{4\br}{d} \right)-\bar f_R\right]
\left( \lich_2 -\frac{2\br}{d}\right), \\
H^{\sigma\sigma} &=& \frac12 \left(\frac{d-1}{d}\right)^2 
\left[ 
P \left(\lich_0 -\frac{\br}{d-1} \right) 
 +Q \right]  
\lich_0^2 \left(\lich_0-\frac{\br}{d-1}\right),
\eea

In the definition of the path integral one has to factor the
volume of the gauge group $\sdiff$.
The standard procedure is the Faddeev-Popov construction.
However, we will follow a somewhat more elegant 
(and ultimately equivalent) procedure
that has been discussed for GR in \cite{Mottola:1995sj},
and adapted to the unimodular case in \cite{Percacci:2017fsy}.
In this approach, the gauge degrees of freedom are isolated
by the York decomposition, and the ghosts emerge from the
Jacobian of the change of variables.
The path integral of the unimodular theory at one loop is
\be
Z=\int(dh^{TT}d\xi d\sigma)J_1 e^{-S^{(2)}} ,
\ee
where $J_1$ is the Jacobian (\ref{conjac}). From the transformation properties of $\xi_\mu$
under an infinitesimal special diffeomorphism
$\epsilon^T_\mu$,
\be
\delta\xi_\mu=\epsilon^T_\mu\ ,
\ee
we see that we can identify $\xi_\mu$ as the coordinate
along the orbits of the gauge group in the space of metrics.
Thus in the path integral we can replace $(d\xi)$
with the integral over the gauge group $(d\epsilon^T)$.

The determinants are (\ref{conTT}), coming from $h_{\mu\nu}^{TT}$,
\bea
\Det \left(\lich_0-\frac{\br}{d-1}\right)^{-1/2}
\Det \left( \lich_0 -\frac{\br}{d-1}
+ \frac{Q}{P}\right)^{-1/2}
\Det\lich_0^{-1} ,
\eea
coming from the scalar $\sigma$,
and finally the Jacobian (\ref{conjac}).
Putting them all together one finds 
\be
Z=\int d\epsilon^T
\frac{\Det \left(\lich_1-\frac{2\br}{d}\right)^{1/2}}{\Det\left(\lich_2-\frac{4\br}{d}-\frac{\bar f_R}{\bar f_X} \right)^{1/2}
\Det\left(\lich_2-\frac{2\br}{d} \right)^{1/2}
\Det \left( \lich_0 -\frac{\br}{d-1}
+ \frac{Q}{P}\right)^{1/2}
\Det\lich_0^{1/2}} .
\ee
Now, as explained in \cite{Percacci:2017fsy},
the invariant measure on the group of special diffeomorphisms
is
$$
\int d\epsilon^T \Det\lich_0^{-1/2}\ ,
$$
so that the last determinant in the denominator
gets absorbed and then factored out
in the overall volume of the gauge group.
Alternatively, one can reach the same conclusion
via the standard Faddeev-Popov procedure,
which has been described for the unimodular case
in \cite{deleon}.

The (Euclidean) effective action is $-\log Z$
and we see that it coincides with the result (\ref{ea}). From here,
the calculation of the divergences proceeds in the same way, leading to (\ref{gammaabc}).
We have thus shown explicitly that the unimodular
theory has the same effective action as the full theory.

%%%%%%%%%%%%%%%%%%%%%%%%%
\section*{Acknowledgment}

This work was supported in part by the Grant-in-Aid for Scientific Research Fund of the JSPS (C) No. 16K05331.
A.D.P acknowledges funding by the DFG, Grant No. Ei/1037-1.

%%%%%%%%%%%%%%%%%%%%%%%%%%%%%%%%%

\end{document}